\documentclass[12pt, letterpaper]{article}  
\usepackage{geometry}                		
\pdfoutput=1            					
\usepackage{graphicx}
\usepackage{caption}
\usepackage{subcaption}		
\usepackage{changepage}

\newlength{\offsetpage}
\newenvironment{widepage}{\begin{adjustwidth}{-\offsetpage}{-\offsetpage}%
    \addtolength{\textwidth}{2\offsetpage}}%
{\end{adjustwidth}}		
		
\usepackage{amssymb}
\usepackage{amsmath}
\usepackage{textcomp}
\usepackage{textgreek}
\usepackage{nicefrac}
\usepackage{threeparttable}
\usepackage{verbatim}

\usepackage{authblk}
\usepackage{url}
\usepackage{gensymb}

\usepackage{etoolbox}
\apptocmd{\sloppy}{\hbadness 10000\relax}{}{}

\makeatletter
\DeclareRobustCommand*\textsubscript[1]{%
  \@textsubscript{\selectfont#1}}
\def\@textsubscript#1{%
  {\m@th\ensuremath{_{\mbox{\fontsize\sf@size\z@#1}}}}}
\makeatother

\title{The Importance of the Electron Mean Free Path for Superconducting RF Cavities}

\author{J. T. Maniscalco}
\author{D. Gonnella}
\author{M. Liepe}
\affil{CLASSE, Cornell University, Ithaca, NY}
\date{July 5, 2016 \\ (Updated July 26, 2016)}							

\begin{document}

\maketitle

\abstract

Impurity-doping is an exciting new technology in the field of SRF, producing cavities with record-high quality factor $Q_0$ and BCS surface resistance that decreases with increasing RF field. Recent theoretical work has offered a promising explanation for this anti-Q-slope, but the link between the decreasing surface resistance and the short mean free path of doped cavities has remained elusive. In this work we investigate this link, finding that the magnitude of this decrease varies directly with the mean free path: shorter mean free paths correspond with stronger anti-Q-slopes. We draw a theoretical connection between the mean free path and the overheating of the quasiparticles, which leads to the reduction of the anti-Q-slope towards the normal Q-slope of long-mean-free-path cavities. We also investigate the sensitivity of the residual resistance to trapped magnetic flux, a property which is greatly enhanced for doped cavities, and calculate an optimal doping regime for a given amount of trapped flux.

\section*{Introduction}

A well-known property of superconductors is their perfect DC conductivity and their resulting ability to expel ambient magnetic flux, as described by Bardeen, Cooper, and Schreiffer (BCS) in 1957\cite{BCS1,BCS2}. Under radio-frequency (RF) electromagnetic fields, the finite inertia of the Cooper pairs causes the superconductor to develop a non-zero surface resistance, $R_\text{s}$, which can be separated into a superposition of the temperature-dependent ``BCS resistance'' $R_\text{BCS}$ and the temperature-independent ``residual resistance'' $R_\text{0}$, with the relation $R_\text{BCS} + R_0 = R_\text{s}$\cite{Hasan}. This BCS resistance typically increases with the strength of the applied field in traditionally prepared (non-doped) cavities. However, under certain circumstances, $R_\text{BCS}$ has been observed to decrease with increasing field strength, an effect often called the ``anti-Q-slope''\footnote{The field-dependent reduction in surface resistance manifests as an increase in the quality factor $Q_0$ of SRF cavities; the effect is often referred to as the ``anti-Q-slope'', to differentiate it from the typically negative ({\em i.e.} increasing resistance) Q-slopes in the medium- and high-field regions.}. In the context of SRF, this phenomenon was discovered in 2013 by A. Grassellino {\em et al.} at Fermilab in niobium doped with nitrogen and by Dhakal {\em et al.} at Jefferson Laboratory\cite{fermi2013,jlab2013}. Impurity-doping SRF cavities in this way causes both this reverse dependence of $R_\text{BCS}$ on the field strength and a decrease in the low-field surface resistance, making the technology highly appealing for new accelerators requiring very efficient SRF cavities.

Unfortunately, doping cavities with nitrogen leads to an increased susceptibility to RF losses from trapped magnetic flux\cite{gonnellaJAP}. Magnetic flux is usually trapped during the cool-down or during quench events by Meissner-effect pinning at impurities and defects. Trapping flux increases the effective residual resistance $R_\text{0}$ of the superconductors, which can be detrimental to accelerator operation. Further, nitrogen-doping has been shown to lower the average quench field of SRF cavities, limiting the peak available accelerating gradient\cite{improvedNdoping}.

In this work we investigate these effects of impurity doping on the surface resistance to find an optimal doping level, as measured by the electron mean free path $\ell$, for niobium SRF accelerating structures. We offer new insight into how the mean free path affects the field dependence of the BCS resistance.

\section*{Field Dependence of the BCS Resistance}

\subsection*{Background}

Though the dependence of the low-field surface resistance on the mean free path observed in superconductors with minimum near ${\ell\approx\xi_0/2}$ is predicted by BCS theory, the dependence of this resistance on the strength of the RF field is not yet fully understood. Recent theoretical work by A. Gurevich offers a promising explanation for the decrease of $R_\text{BCS}$ with increasing RF field\cite{gurevich2014}. Here, the RF field changes the density of states of the Bogoliubov quasiparticles in a way that tends to reduce their number density and thus reduce the Ohmic losses due to their movement. An important effect in this theory is the overheating of the quasiparticles, which offers an opposing force to the reduction of the surface resistance by increasing the effective temperature of the quasiparticles. This overheating is controlled by the ``overheating parameter''~$\alpha$, encapsulating material parameters such as the Kapitza interface conductance $h_\text{K}$, the thermal conductivity $\kappa$, and $Y,$ the energy transfer rate between quasiparticles and phonons, by the relations\footnote{Equations \ref{eq:g_13} and \ref{eq:g_14} are Eqs. 13 and 14, respectively, in \cite{gurevich2014}.} in Eqs. \ref{eq:g_13} and \ref{eq:g_14}:

\begin{align}
T-T_0 &= \frac{\alpha T_0}{R_\text{s,0}}\left(\frac{H_\text{a}}{H_\text{c}}\right)^2 R_\text{s}\left(H_\text{a}, T\right) \label{eq:g_13} \\
\alpha &= \frac{R_\text{s,0}B_\text{c}^2}{2\mu_0^2T_0}\left(\frac{1}{Y} + \frac{d}{\kappa} + \frac{1}{h_K}\right) \label{eq:g_14}
\end{align}

Here, $T$ is the quasiparticle temperature, $T_0$ is the experimental bath temperature, $R_\text{s,0}$ is the low-field BCS resistance, $\mu_0H_\text{a} = B_\text{a}$ is the RF magnetic field at the surface, $\mu_0H_\text{c} = B_\text{c}$ is the thermodynamic critical field, $R_\text{s}$ is the BCS resistance at the given applied field and temperature, and $d$ is the thickness of the SRF cavity wall.

\subsection*{Experimental Procedures}

In order to study the effects of nitrogen doping in niobium SRF cavities, we prepared many cavities using nitrogen-doping techniques to achieve a range of values of the mean free path $\ell$ in the RF penetration layer, ranging from 4~nm to over 200~nm\footnote{It is necessary to mention here that the doped layer is typically 5-50~\textmu m thick and does not extend through the bulk; $\ell$ is essentially uniform over the RF penetration depth, which is on the order of 100~nm. This is not true for 120\degree C-baked cavities, which are not discussed in this work.}. These preparations were done on single-cell 1.3~GHz TESLA-shape cavities\cite{TESLA}. Table \ref{tab1} summarizes their properties and preparation techniques. We performed vertical RF tests of these cavities, measuring the quality factor $Q_0$ as a function of field at many temperatures, as well as the low-field $Q_0$ and resonant frequency as a function of temperature\cite{gonnelladependence,gonnellaflux,improvedNdoping,gonnellathesis}. From these, for each cavity test we extracted the residual resistance as a function of field, the BCS resistance as a function of field and temperature, the energy gap $\Delta$, the mean free path $\ell$, penetration depth $\lambda$ as a function of temperature, coherence length $\xi$, and critical temperature $T_\text{c}$, using the methods described in \cite{meyers,vallesthesis,posenthesis,gonnellathesis}. We then used these parameters to calculate theoretical predictions from \cite{gurevich2014}, using the overheating parameter $\alpha$ as a fitting parameter. We found it necessary to use an additional field-independent fitting parameter, $s$, as a scaling factor, with the relation $s = R_\text{BCS,meas}/R_\text{thy}$. For a given cavity, $\alpha$ was allowed to vary for all temperatures, while all temperatures shared a single value of $s$. The scaling factor $s$ was typically near unity; we believe that this accounted for systematic experimental errors, which we usually cite as 10\%. We discuss this further below. It is important to note here that the theory does not include any explicit dependence of the field-dependent resistance on the mean free path; in our analysis, we sought to investigate any possible dependence of the overheating parameter on $\ell$.

\begin{table*}[t]
\begin{widepage}
\centering
\begin{threeparttable}
\caption{Overview of Cavity Preparations\cite{gonnellaJAP}}
\begin{tabular}{ccccc} \hline \hline
Cavity & Preparation & $T_c$ [K] & $\Delta/k_BT_c$ & Mean Free Path \\
\hline
C3(P2) & 990$^\circ$C N-doping\tnote{1} + 5 \textmu m VEP & $9.1\pm0.1$ & $2.05\pm0.01$ & $4\pm1$ \\
C2(P2) & 900$^\circ$C N-doping\tnote{2} + 18 \textmu m VEP  & $9.1\pm0.1$ & $2.00\pm0.01$ & $6\pm1$  \\
C2(P3) & 900$^\circ$C N-doping\tnote{2} + 6 \textmu m VEP  & $9.2\pm0.1$ & $1.94\pm0.01$ & $17\pm5$ \\
C2(P1) & 800$^\circ$C N-doping\tnote{3} + 6 \textmu m VEP& $9.3\pm0.1$ & $1.88\pm0.01$ & $19\pm6$ \\ 
C3(P1) & 800$^\circ$C N-doping\tnote{3} + 12 \textmu m VEP& $9.3\pm0.1$ & $1.91\pm0.01$ & $34\pm10$\\
C1(P1) & 800$^\circ$C N-doping\tnote{3} + 18 \textmu m VEP& $9.3\pm0.1$ & $1.88\pm0.01$ & $39\pm12$ \\
C4(P1) & 800$^\circ$C N-doping\tnote{3} + 24 \textmu m VEP& $9.2\pm0.1$ & $1.89\pm0.01$ & $47\pm14$ \\
C5(P1) & 800$^\circ$C N-doping\tnote{3} + 30 \textmu m VEP& $9.2\pm0.1$ & $1.88\pm0.01$ & $60\pm18$ \\
C5(P2) & 800$^\circ$C N-doping\tnote{3} + 40 \textmu m VEP& $9.2\pm0.1$ & $1.94\pm0.01$ & $213\pm64$ \\
\hline \end{tabular}
\begin{tablenotes}
\item[1] 100 $\mu$m VEP, 800$^\circ$C in vacuum for 3 hours, 990$^\circ$C in 30 mTorr of N$_2$ for 5 minutes, 5 $\mu$m VEP.
\item[2] 100 $\mu$m VEP, 800$^\circ$C in vacuum for 3 hours, 900$^\circ$C in 60 mTorr of N$_2$ for 20 minutes, 900$^\circ$C in vacuum for 30 minutes, light EP.
\item[3] 100 $\mu$m VEP, 800$^\circ$C in vacuum for 3 hours, 800$^\circ$C in 60 mTorr of N$_2$ for 20 minutes, 800$^\circ$C in vacuum for 30 minutes.
\end{tablenotes}
\label{tab1}
\end{threeparttable}
\end{widepage}
\end{table*}

\subsection*{$R_\text{BCS}$ Results and Theoretical Fits}

\begin{figure}[p]
\begin{widepage}
\centering
    \begin{subfigure}[t]{0.31\textwidth}
    \centering
    \includegraphics[width=\linewidth]{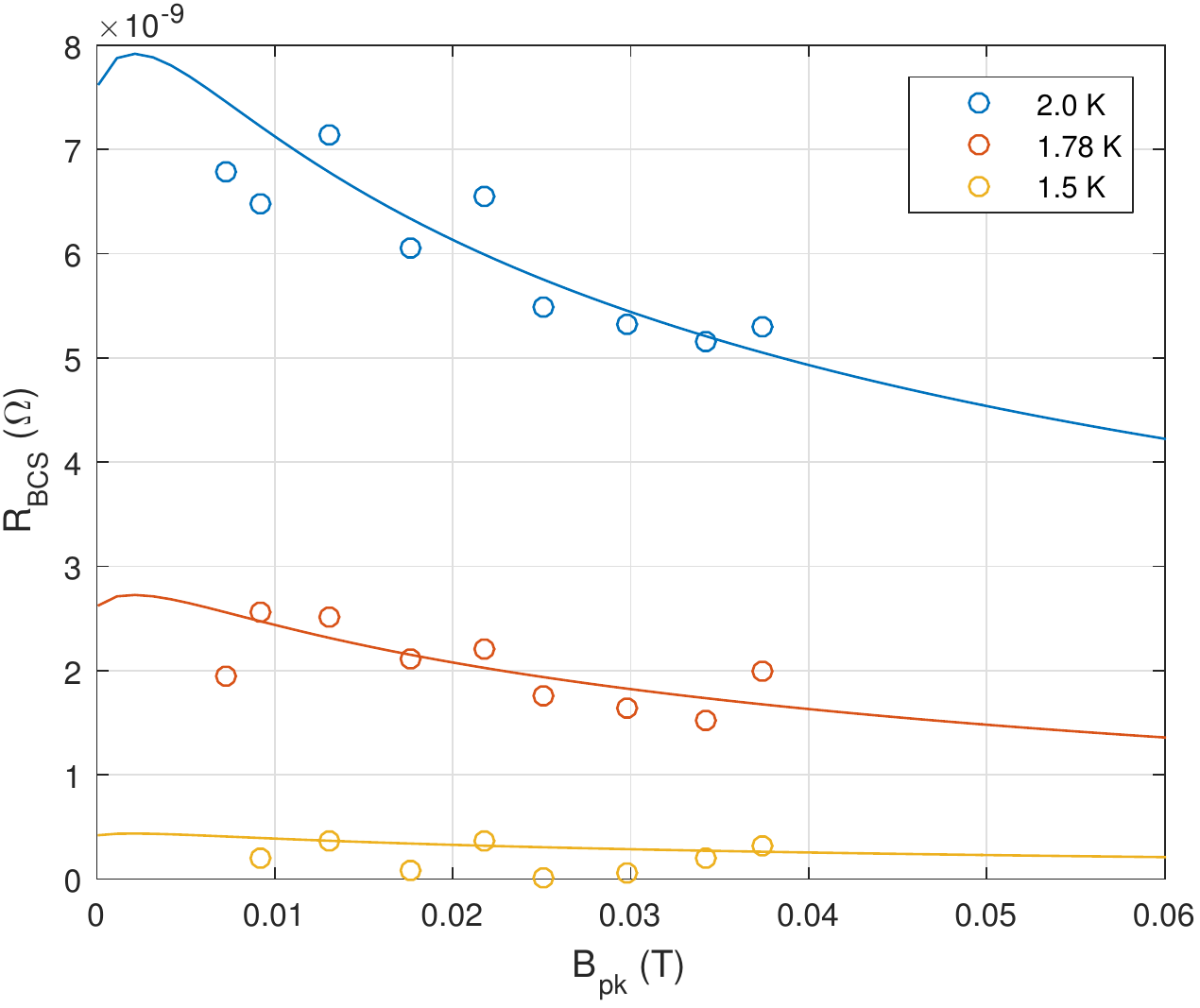}
    \caption{Test of cavity C3(P2), with $\ell~=~4.5\pm1.3$~nm, $\alpha \approx 0$. \label{fig:te13_overdope}}
    \end{subfigure}\hfill
    \begin{subfigure}[t]{0.31\textwidth}
    \centering
    \includegraphics[width=\linewidth]{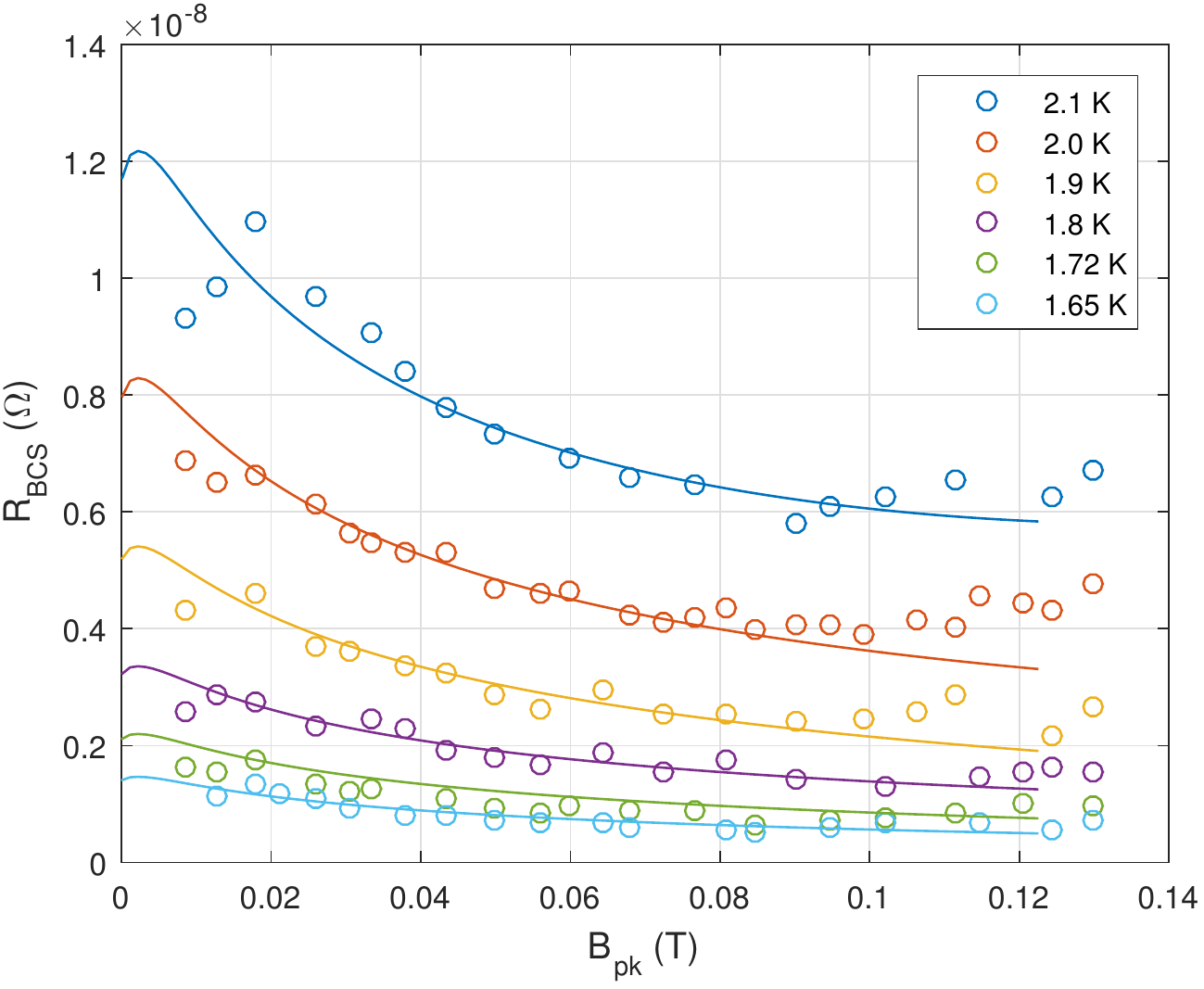}
    \caption{Test of cavity C3(P1), with $\ell~=~34\pm10$~nm, $\alpha(2.1~K) = 0.44$. \label{fig:te13_mediumdope}}
    \end{subfigure}\hfill
    \begin{subfigure}[t]{0.31\textwidth}
    \centering
    \includegraphics[width=\linewidth]{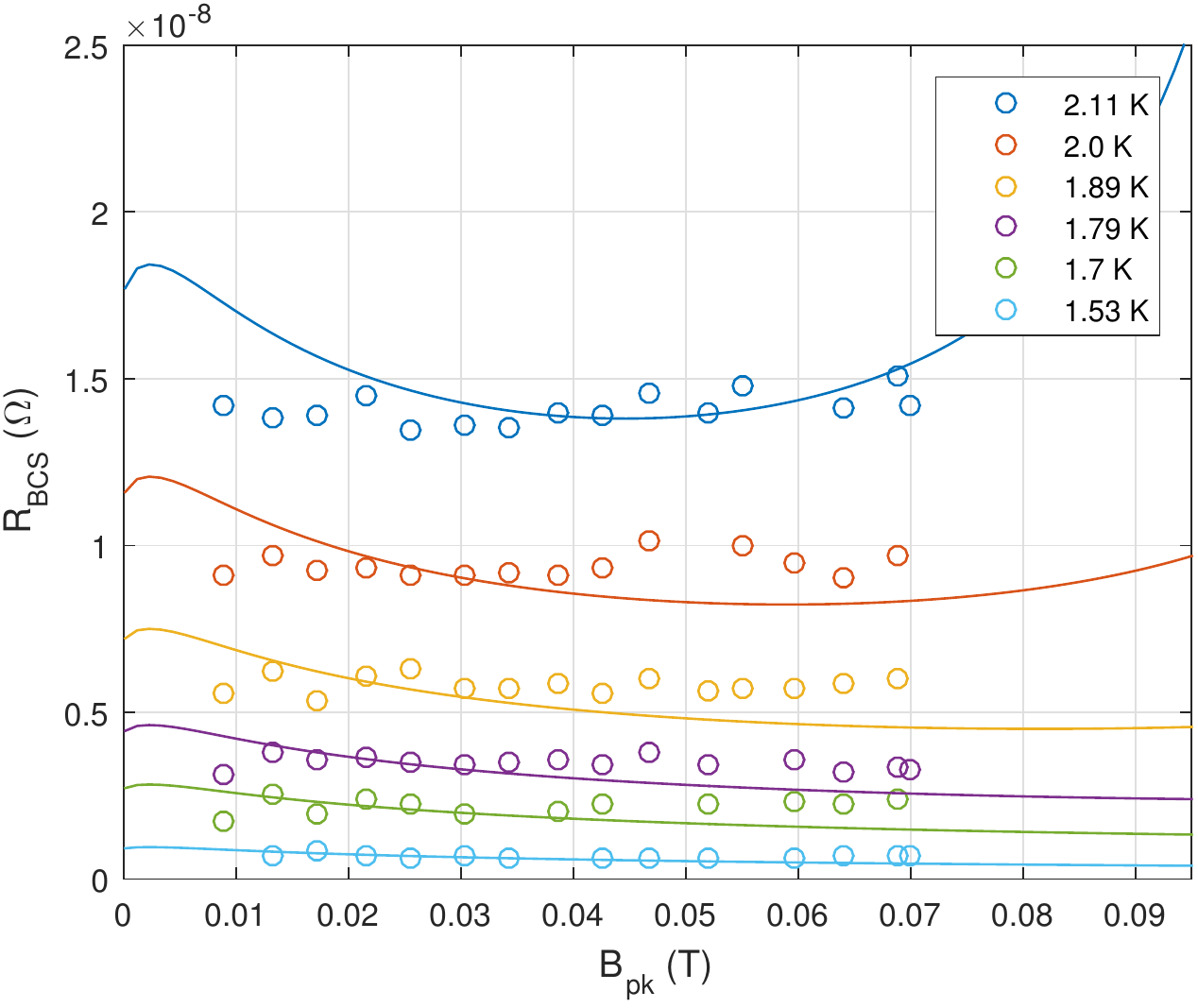}
    \caption{Test of cavity C5(P1), with $\ell~=~60\pm18$~nm, $\alpha(2.11~K) = 1.7$. \label{fig:te15}}
    \end{subfigure}
\end{widepage}
\caption{$R_\text{BCS}$ as a function of peak surface magnetic field for several representative 1.3~GHz TESLA cavities of varying mean free path $\ell$.  Lines show the fitted theoretical predictions with given overheating parameter $\alpha$. Error bars have been omitted for visual clarity, but are typically cited at 10\%, largely due to systematic errors that have been accounted for by the scaling parameter $s$.\label{fig:R_vs_B}}
\end{figure}

Figure \ref{fig:R_vs_B} shows typical results of the BCS surface resistance as a function of applied RF magnetic field, as well as the theoretical predictions based on \cite{gurevich2014}. Figure \ref{fig:te13_overdope} shows the measurements of the BCS resistance for a heavily-doped cavity, with mean free path of $4.5\pm1.3$~nm, achieved with a 990\degree C nitrogen bake followed by chemical removal (etching) of 5~\textmu m of material. Here, the theoretical predictions match very well with experimental results, with no quasiparticle overheating ($\alpha \approx 0)$\footnote{By the theory, $\alpha$ should never get to zero due to the contributions from $\kappa$ and $h_\text{K}$, but the effects of these on $R_\text{BCS}$ in this case are minimal\cite{gurevich2014}.}. Figure \ref{fig:te13_mediumdope} shows similar measurements for a cavity with medium doping level, with mean free path of $34\pm10$~nm after an 800\degree C nitrogen bake and a 12~\textmu m etch. Again, there is good agreement with theory and experiment. Here an overheating parameter of $\alpha = 0.44$ was found to best fit the results of the RF measurements at 2.1~K, with $\alpha$ generally decreasing with decreasing temperature. Figure \ref{fig:te15} shows the results of a lightly doped cavity, with mean free path of $60\pm18$~nm after receiving an 800\degree C nitrogen bake followed by 30~\textmu m of material removal. Here, with $\ell$ approaching the clean limit, experimental results and theoretical predictions diverge; though the theoretical predictions are well within the quoted uncertainty interval, we believe that this uncertainty accounts mainly for systematic errors that are absorbed into the scaling parameter $s$. The divergence in the overall trends of the theoretical predictions and the experimental data suggest that the thermal effects become characteristically different at higher mean free path, as the anti-Q-slope transitions to the more commonly observed medium-field Q-slope, suggesting nonlinear effects.

\begin{figure}[p]
\centering
\includegraphics[height=0.45\textwidth]{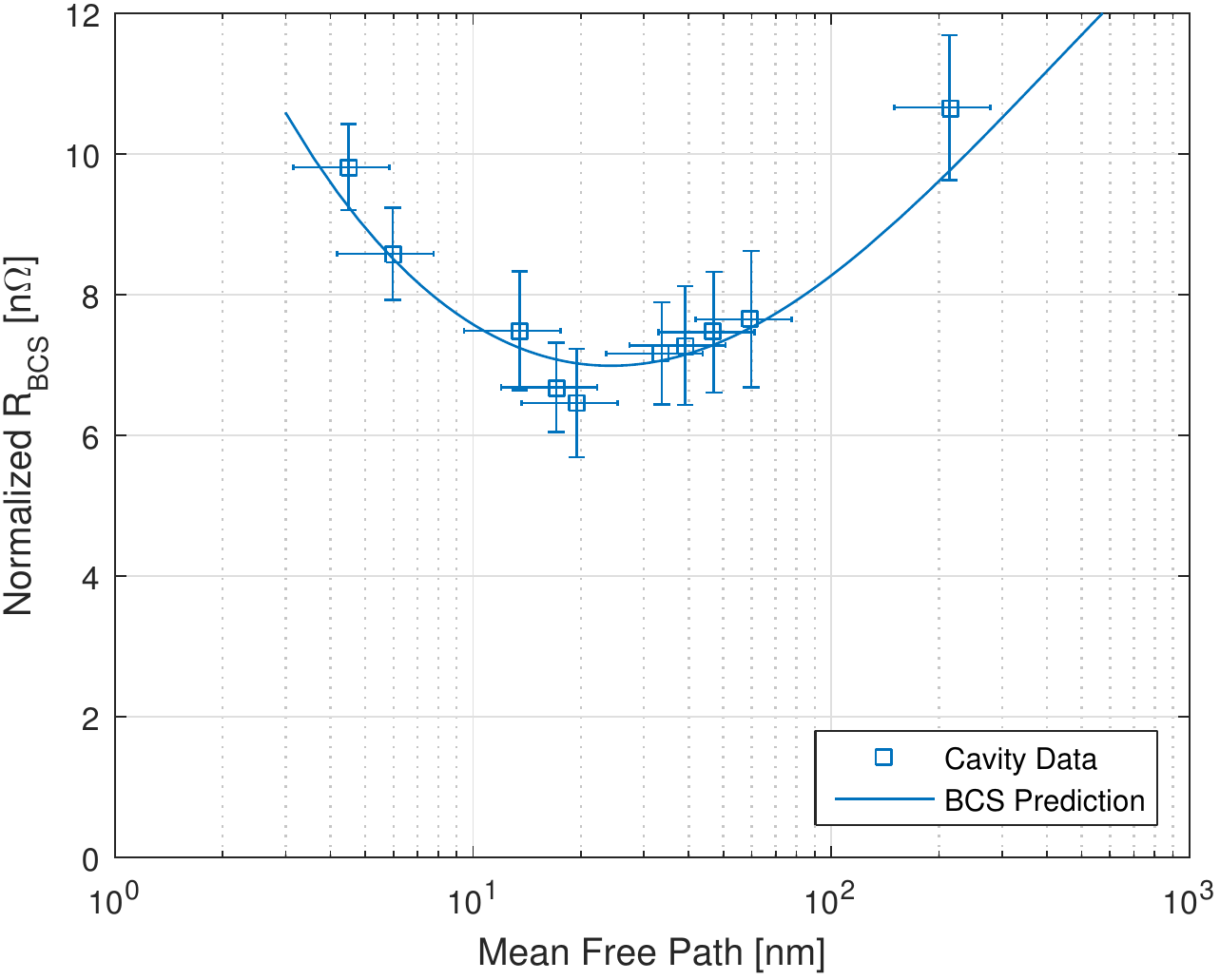}
\caption{At low field ($\sim5$~MV/m), the measured BCS resistance (normalized to an average energy gap $\Delta$ and critical temperature $T_\text{c}$) at 1.3~GHz and 2~K, plotted as a function of electron mean free path, shows an agreement with the theoretical minimum of $25\pm1$~nm. \label{fig:rbcs_5_vs_mfp}}
\end{figure}

Figure \ref{fig:rbcs_5_vs_mfp} shows the low-field 2~K BCS resistance\footnote{The BCS resistance here, and later at 16~MV/m, has been normalized to average critical temperature and energy gap, by the formula $R_\text{BCS, norm} = R_\text{BCS}\exp\left(-\Delta_\text{avg}/kT_\text{c,avg} + \Delta/kT_\text{c}\right)$. We do not see any strong dependence of the energy gap on doping level, suggesting that this is a valid normalization technique.} of many of these tests, measured at $\sim20$~mT (corresponding to an accelerating gradient of $\sim5$~MV/m in a TESLA cavity), as a function of the mean free path $\ell$. This resistance corresponds very well with the theoretical 2~K BCS resistance, also shown in Fig. \ref{fig:rbcs_5_vs_mfp}, which has its minimum at $\ell=25\pm1$~nm for average material parameters. This is a good confirmation of the accuracy of our measurement techniques.

\subsection*{Quasiparticle Overheating and the Mean Free Path}

\begin{figure}[p]
\begin{widepage}
\centering
    \begin{subfigure}[t]{0.48\textwidth}
    \centering
    \includegraphics[width=\linewidth]{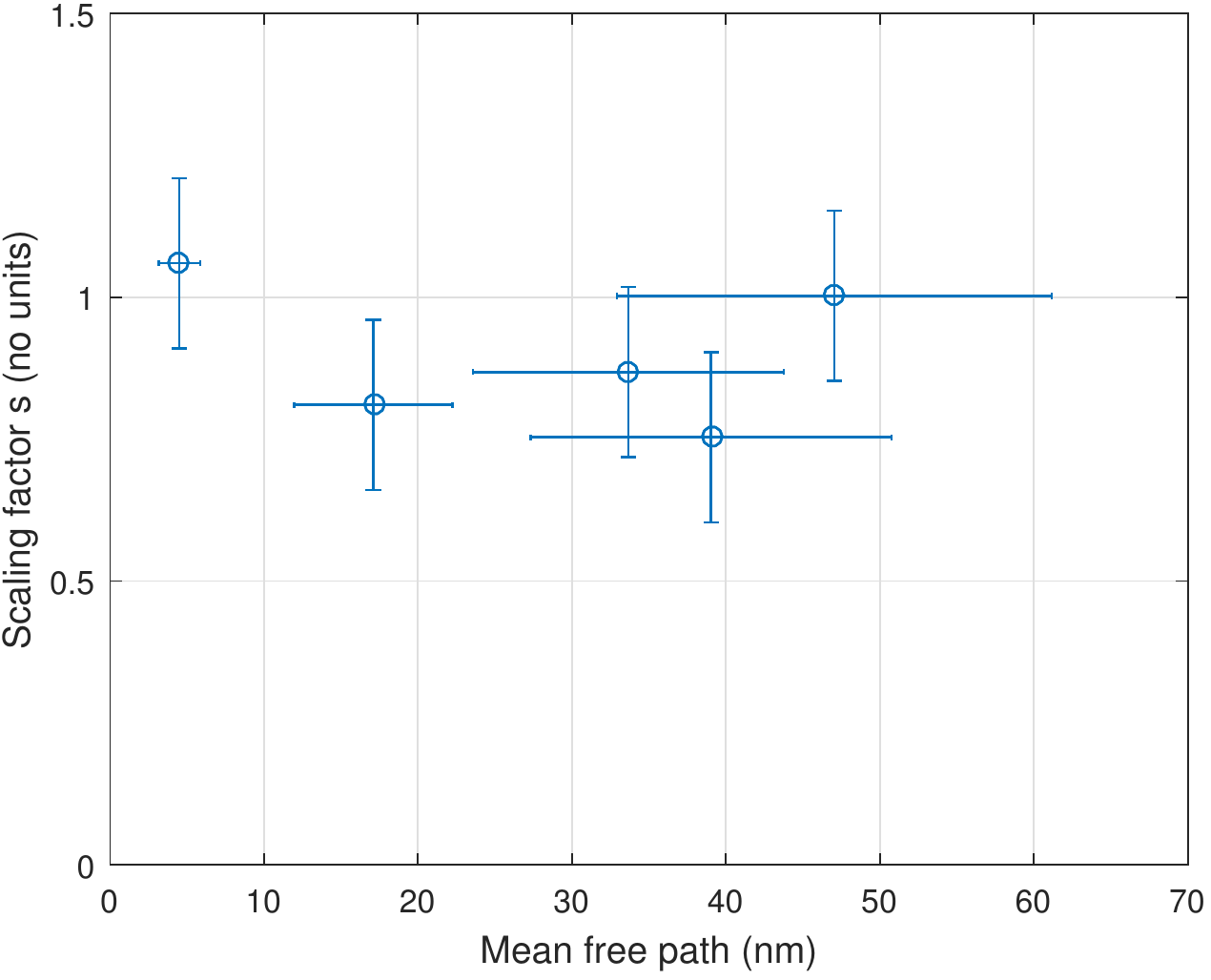}
    \caption{Scaling factor $s$ plotted over mean free path. \label{fig:s_vs_mfp}}
    \end{subfigure}\hfill
    \begin{subfigure}[t]{0.48\textwidth}
    \centering
    \includegraphics[width=\linewidth]{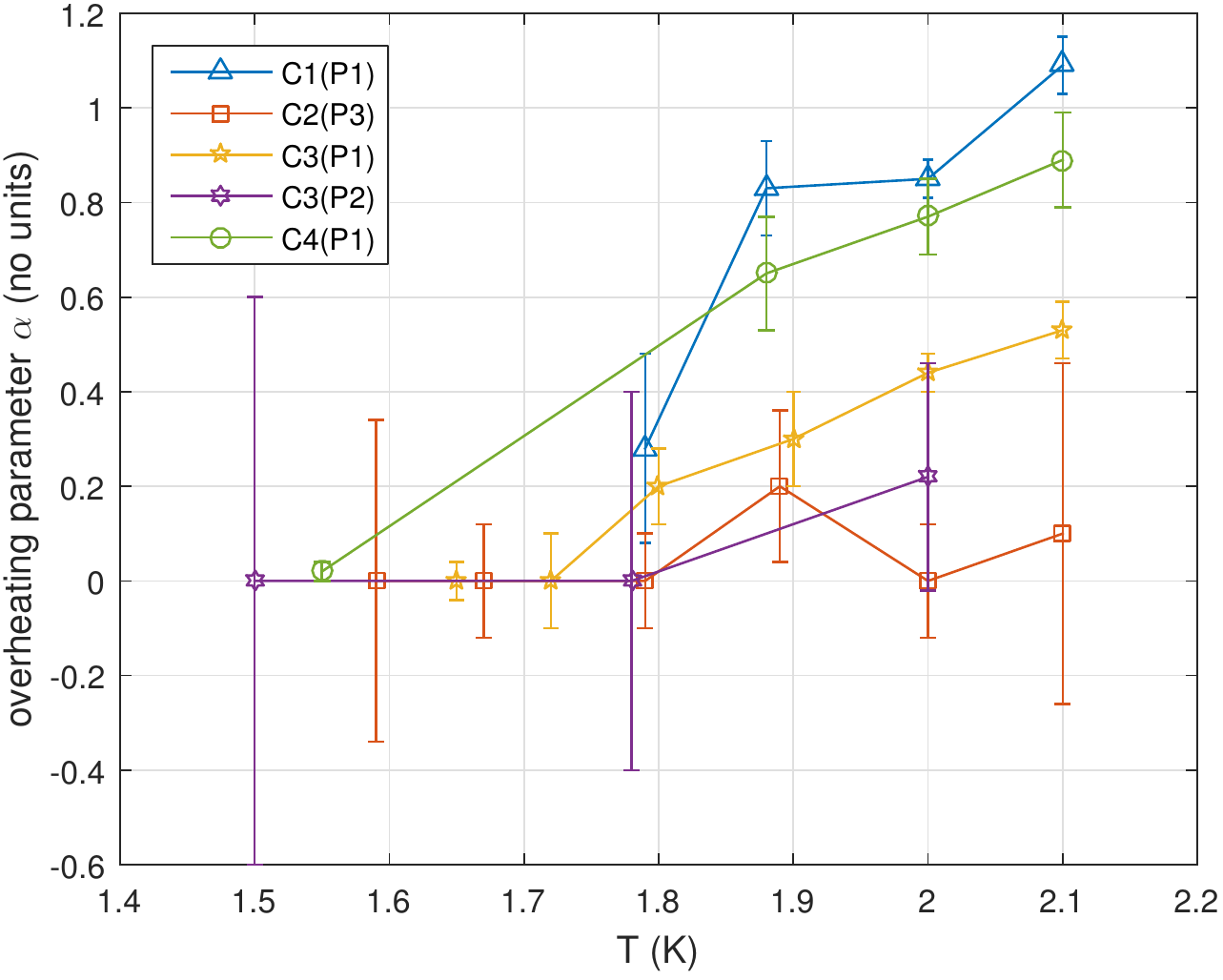}
    \caption{Overheating parameter $\alpha$ plotted over temperature for many cavities. \label{fig:a_vs_T}}
    \end{subfigure}

    \begin{subfigure}[t]{0.48\textwidth}
    \centering
    \includegraphics[width=\linewidth]{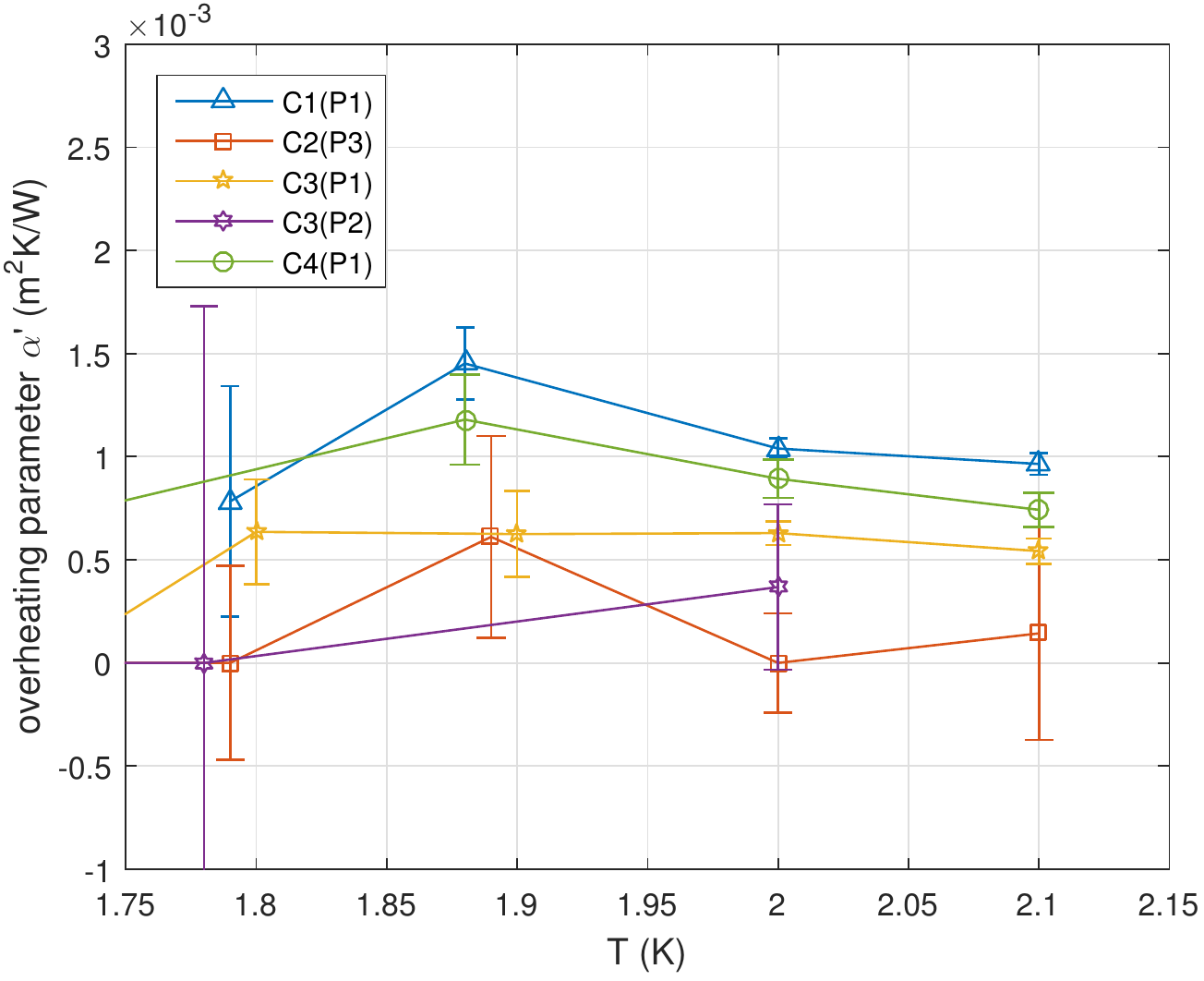}
    \caption{Normalized overheating parameter $\alpha'$ plotted over temperature for many cavities.\label{fig:a_normalized_vs_T}}
    \end{subfigure}\hfill
    \begin{subfigure}[t]{0.48\textwidth}
    \centering
    \includegraphics[width=\linewidth]{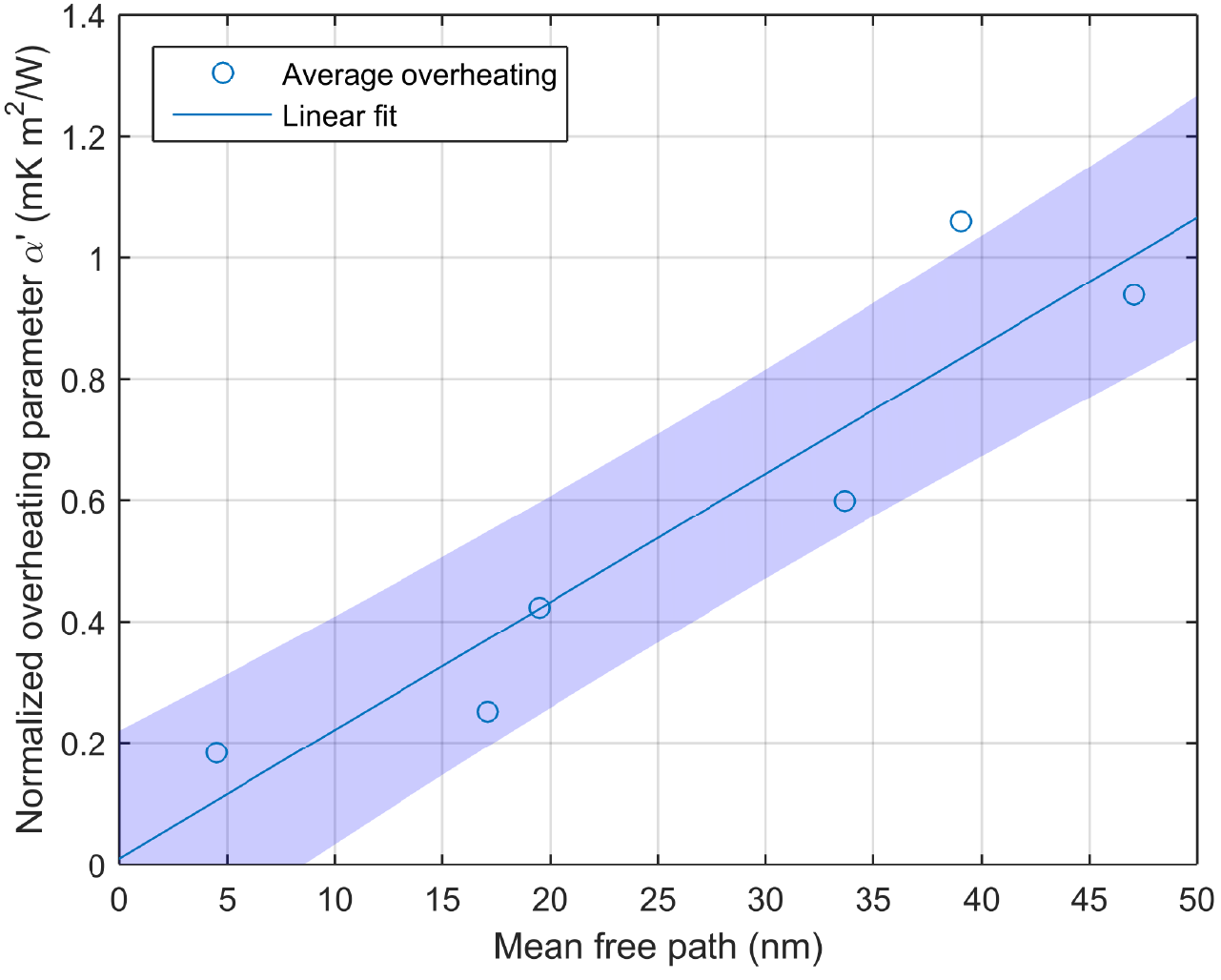}
    \caption{Average normalized overheating parameter $\alpha'$ plotted as a function of mean free path, with a linear fit Shaded region indicates 1$\sigma$ confidence interval. \label{fig:a_normalized_vs_mfp}}
    \end{subfigure}
\end{widepage}
\caption{Theory fit parameters over varying mean free path and experimental temperature.}
\end{figure}

We next investigated the two fitting parameters, $s$ and $\alpha$, for dependences on any material or superconductor parameters. The scaling parameter $s$, which was fixed for each cavity at all temperatures, shows little dependence on any physical parameters. The dependence on mean free path, or lack thereof, is shown in Fig. \ref{fig:s_vs_mfp}. For $\alpha$, we saw a general positive slope with increasing temperature, as mentioned above and as shown in Figure \ref{fig:a_vs_T}. This trend is not surprising given the dependence of $\alpha$ on the BCS surface resistance (see Eq. \ref{eq:g_14}), which increases rapidly with temperature. In an effort to account for the dependence of the quasiparticle overheating on the experimental temperature and focus our study on the dependence on actual thermal parameters, we derive from the overheating parameter $\alpha$ a normalized overheating parameter $\alpha'$ by Eq. \ref{eq:a_prime}:

\begin{equation}
\alpha' = \alpha\frac{2\mu_0^2T_0}{R_\text{s0}B_\text{c}^2} \label{eq:a_prime}
\end{equation}

This reduces Eqs. \ref{eq:g_13} and \ref{eq:g_14} to Eqs. \ref{eq:g13} and \ref{eq:g14}:

\begin{align}
T - T_0 &= \frac{1}{2}\alpha'H_\text{a}^2R_\text{s}(H_\text{a}, T)\label{eq:g13}\\
\alpha' &= \left(\frac{1}{Y} + \frac{d}{\kappa} + \frac{1}{h_\text{K}}\right)\label{eq:g14}
\end{align}

When plotting this $\alpha'$ as a function of temperature, we see that the dependence has been removed to first order, as shown in Fig. \ref{fig:a_normalized_vs_T}. This suggests that the quasiparticle overheating is independent of temperature to first order, at least within this band of temperatures between 1.8~K and 2.1~K for the cases where $\alpha'$ is large enough compared to its uncertainty to extract relative trends with $T_0$.

However, plotting as a function of mean free path, we see a clear dependence that is well approximated by a linear relationship with an offset at zero. Figure \ref{fig:a_normalized_vs_mfp} shows this result, with an offset linear fit for $\alpha'$ averaged over temperature. Averaging the values over all temperatures, given the temperature-independence indicated in Fig. \ref{fig:a_normalized_vs_T}, we found an offset of $\gamma = 0.01\pm0.20\times10^{-3}$~K\,m$^2$/W and a slope $\beta = 2.1\pm0.8\times10^4$~K\,m/W, according to the relation in Eq. \ref{eq:a_prime_mfp}:

\begin{equation}
\alpha' = \gamma + \beta\ell \label{eq:a_prime_mfp}
\end{equation}

Quite important to emphasize here is that this approximation is only valid in the short-mean-free-path limit, as the overheating becomes nonlinear for $\ell~>~50$~nm.

This relation makes a compelling suggestion that the mean free path plays a strong, fundamental role in the overheating of the quasiparticles, resulting in the field-dependent surface resistance. Shorter mean free path correlates with smaller overheating and therefore stronger reduction in $R_\text{BCS}$ with increasing RF field. Of note here is that the offset parameter $\gamma$ is on a scale similar to that of the minimal overheating contribution from the Kapitza interface conductance and the thermal conductivity; by Eq. \ref{eq:g14}, for typical values of these properties\cite{gurevich2014}, $(\nicefrac{d}{\kappa} + \nicefrac{1}{h_K})\approx 5\times10^{-4}$~K\,m$^2$/W, on the same order of magnitude as the confidence interval of the linear fit (as seen in Fig. \ref{fig:a_normalized_vs_mfp}). Further, this quantity depends on the properties of the bulk superconductor, not the RF penetration layer; as such, we should not expect it to change significantly with doping level.

This suggests that $(\beta\ell)^{-1}$ is approximately equal to the parameter $Y$ in Eq. \ref{eq:g14}, in this short-mean-free-path limit. In this case, longer mean free path indicates a lower energy transfer rate from quasiparticles to phonons. A potential explanation for this is elastic scattering of the quasiparticles on impurities; quasiparticles moving on lattices with longer mean free path ({\em i.e.} longer spacing between impurity scattering sites) have longer characteristic elastic scattering times and thus lower energy transfer rates, leading to the stronger overheating observed for long mean free path $\ell$.

\begin{figure}[htb]
\centering
\includegraphics[height=0.5\textwidth]{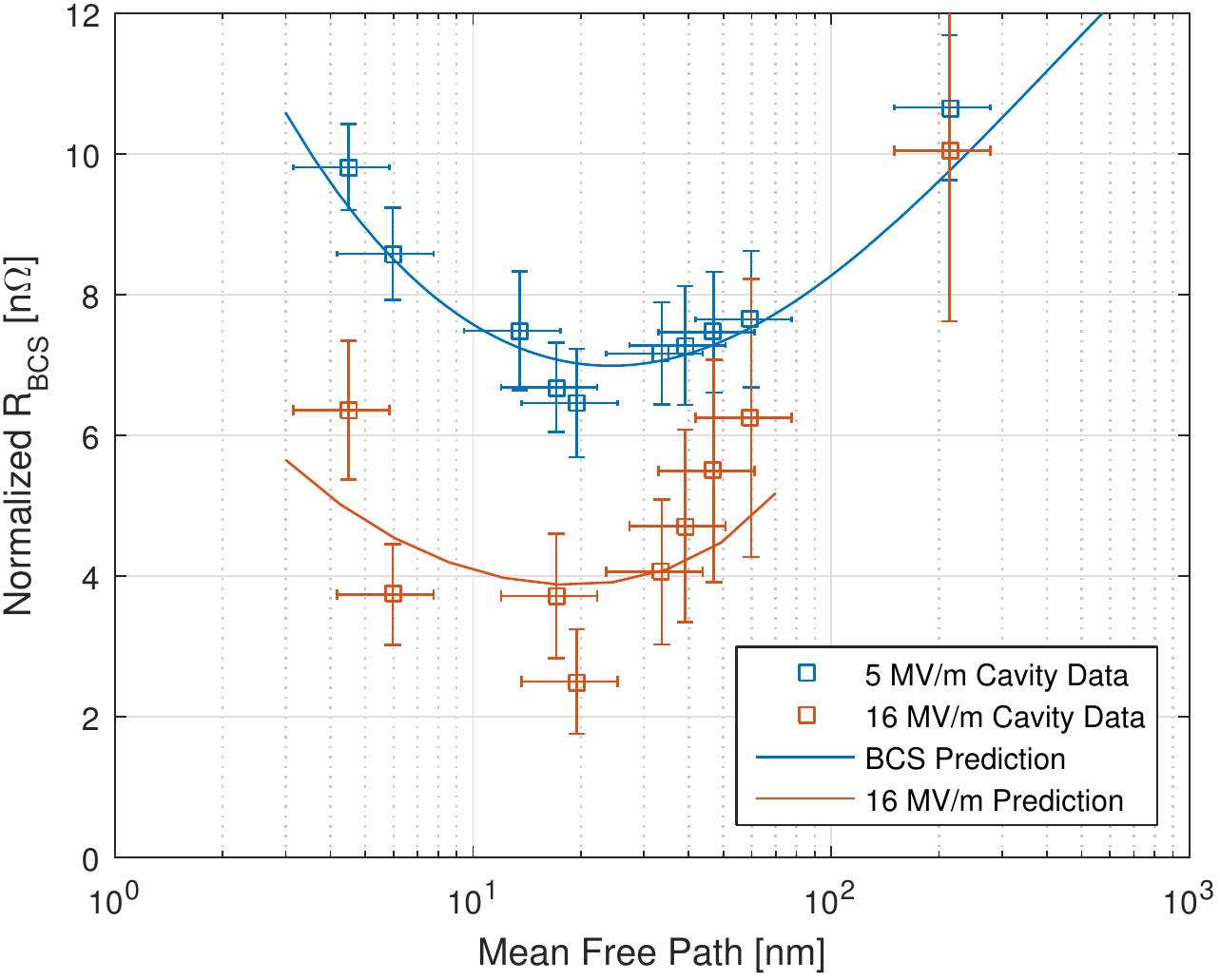}
\caption{BCS resistance at 2~K and 1.3~GHz as a function of mean free path. Blue points and curves are low field data, taken at 5~MV/m (21~mT). Red points are the experimental data at 16~MV/m (68~mT), and the lower red curve is the theoretical BCS resistance extrapolated to 16~MV/m by the modified mean-free-path-dependent theory. Experimental points are normalized to an average $\Delta$ and $T_\text{c}$. Here the 16~MV/m theory curve does not extend beyond $\ell=70$~nm, where the overheating effects become highly nonlinear.\label{fig:rbcs_5_and_16}}
\end{figure}

Using this dependence of $\alpha'$ on $\ell$, we generated theoretical predictions of the field dependence of the surface resistance for a range of mean free paths. We calculated the surface resistance at 68~mT ($E_\text{acc}\approx16$~MV/m for a TESLA cavity), the operating accelerating gradient spec of LCLS-II, to compare to experimental data at the 16~MV/m operating gradient and at low field. Figure \ref{fig:rbcs_5_and_16} shows the low field BCS resistance (normalized in $\Delta$ and $T_\text{c}$, as described above) at 21~mT ($\sim$5~MV/m), the theoretical BCS resistance for average material parameters, the theoretical BCS resistance at 68~mT calculated from \cite{gurevich2014} and the $\alpha'(\ell)$ model, and the experimental data at 68~mT. We see good agreement with the high-field experimental data, and in particular with the shifted minimum in $R_\text{BCS}$ as a function of $\ell$, to about 17~nm at 68~mT from the low-field minimum of 25~nm for these material parameters, a 32\% shift.

\subsection*{Logarithmic Q-Slope Fitting}

In another, parallel analysis, we analyzed the cavity test results empirically to quantify the field-dependent resistance. We found that, in the medium-field region (typically 10-100~mT), the surface resistance followed a logarithmic slope\footnote{Similar results were observed in \cite{gglogslope}.}, following the relation in Eq. \ref{eq:log_slope}:

\begin{align}
R_\text{BCS} = X\log\left(\frac{B_\text{pk}}{[mT]}\right) +Y \label{eq:log_slope}
\end{align}

\begin{figure}[p]
\centering
\includegraphics[height=0.45\textwidth]{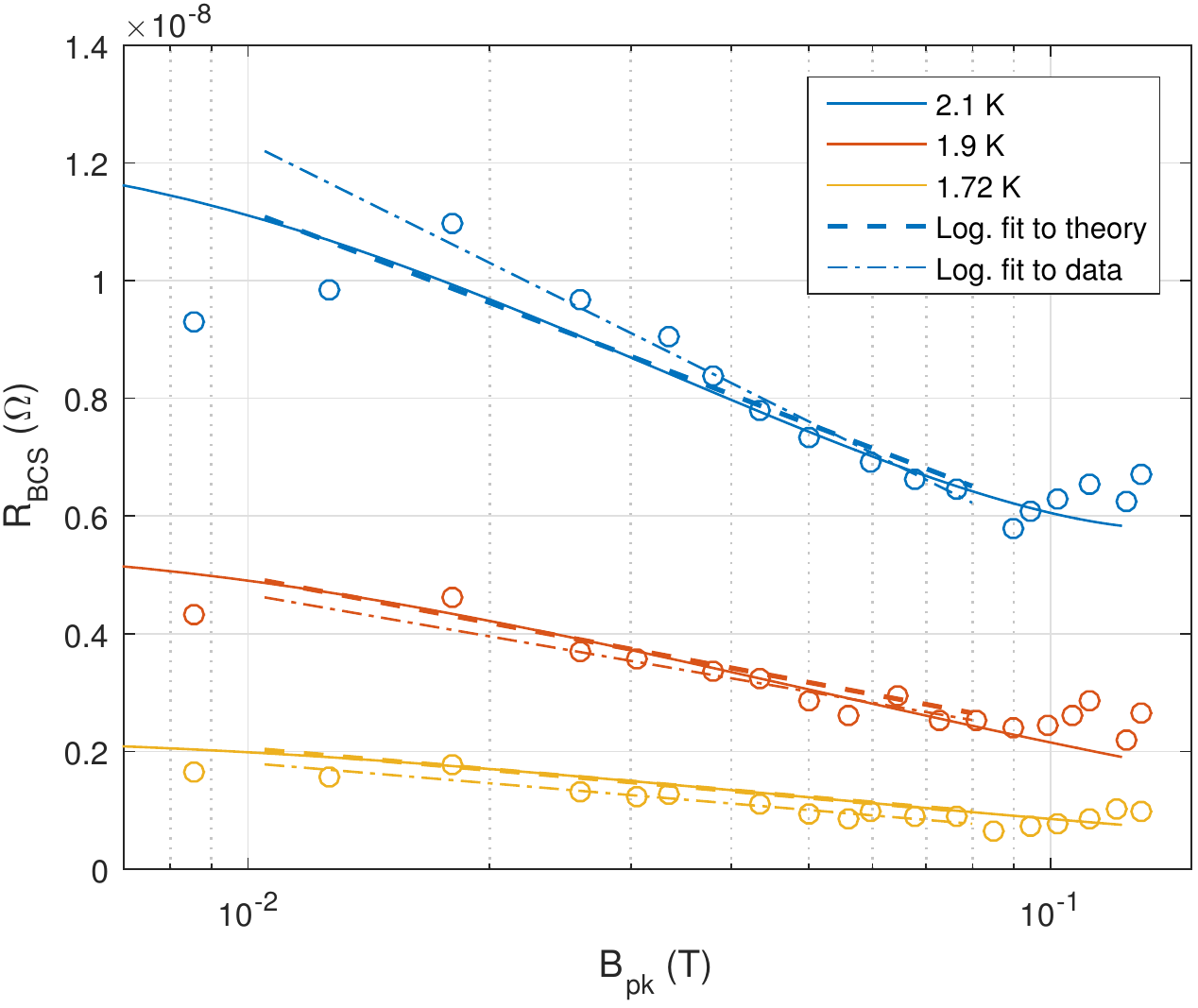}
\caption{C3(P1) results with logarithmic fitting to both experimental data and theoretical predictions. Both show approximate logarithmic behavior in the medium-field region. The plot has been limited to three temperatures to reduce visual clutter.\label{fig:te13_log}}
\end{figure}

\begin{figure}[p]
\begin{widepage}
    \begin{subfigure}[t]{0.48\textwidth}
    \centering
    \includegraphics[width=\linewidth]{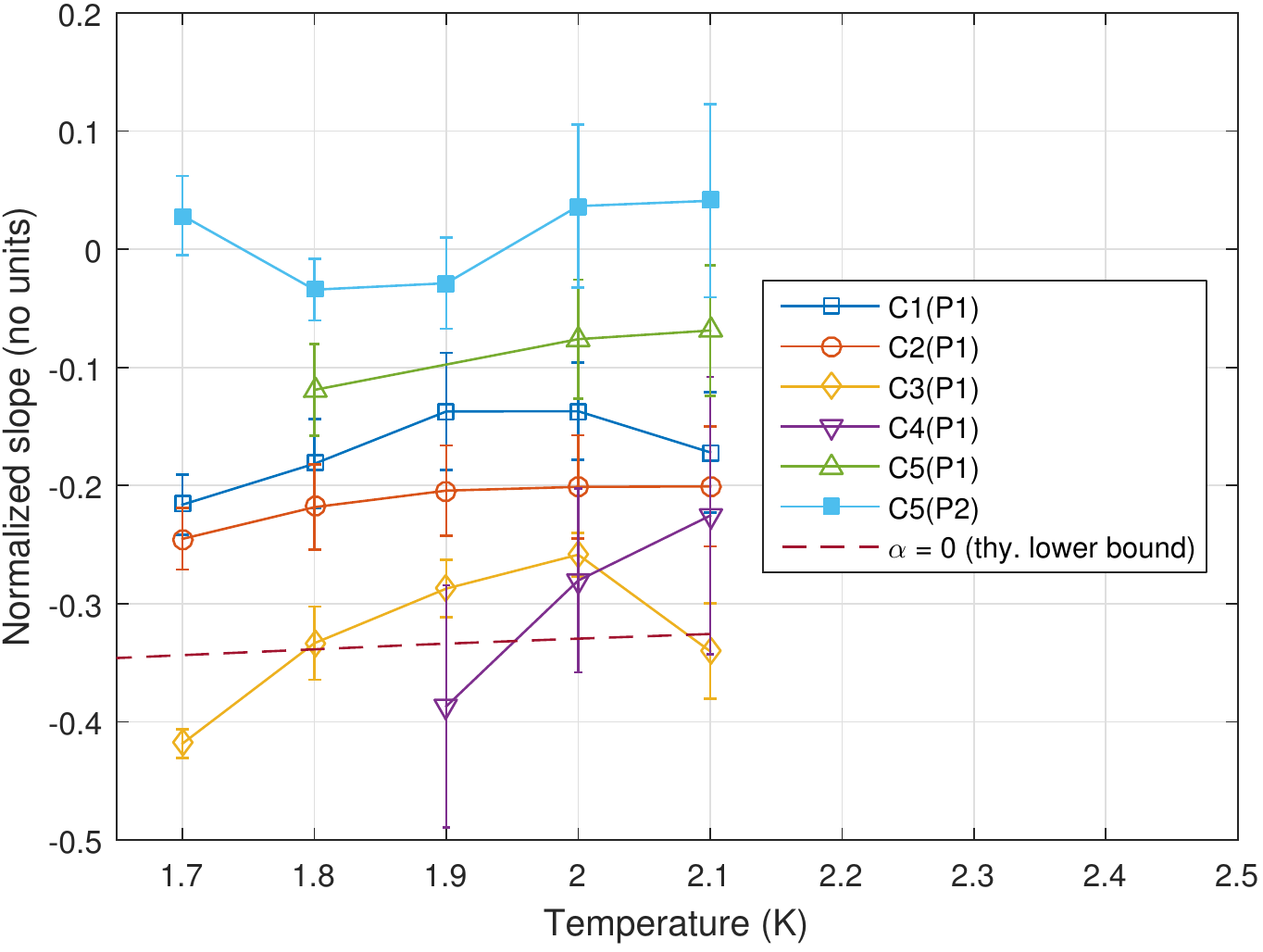}
    \caption{Normalized logarithmic slopes of the surface resistance in field, as defined in Eq. \ref{eq:log_slope}, plotted over temperature. Dashed line indicates the theoretical normalized slope for zero overheating. \label{fig:log_slope_normalized}}
    \end{subfigure}\hfill
    \begin{subfigure}[t]{0.45\textwidth}
    \centering
    \includegraphics[width=\linewidth]{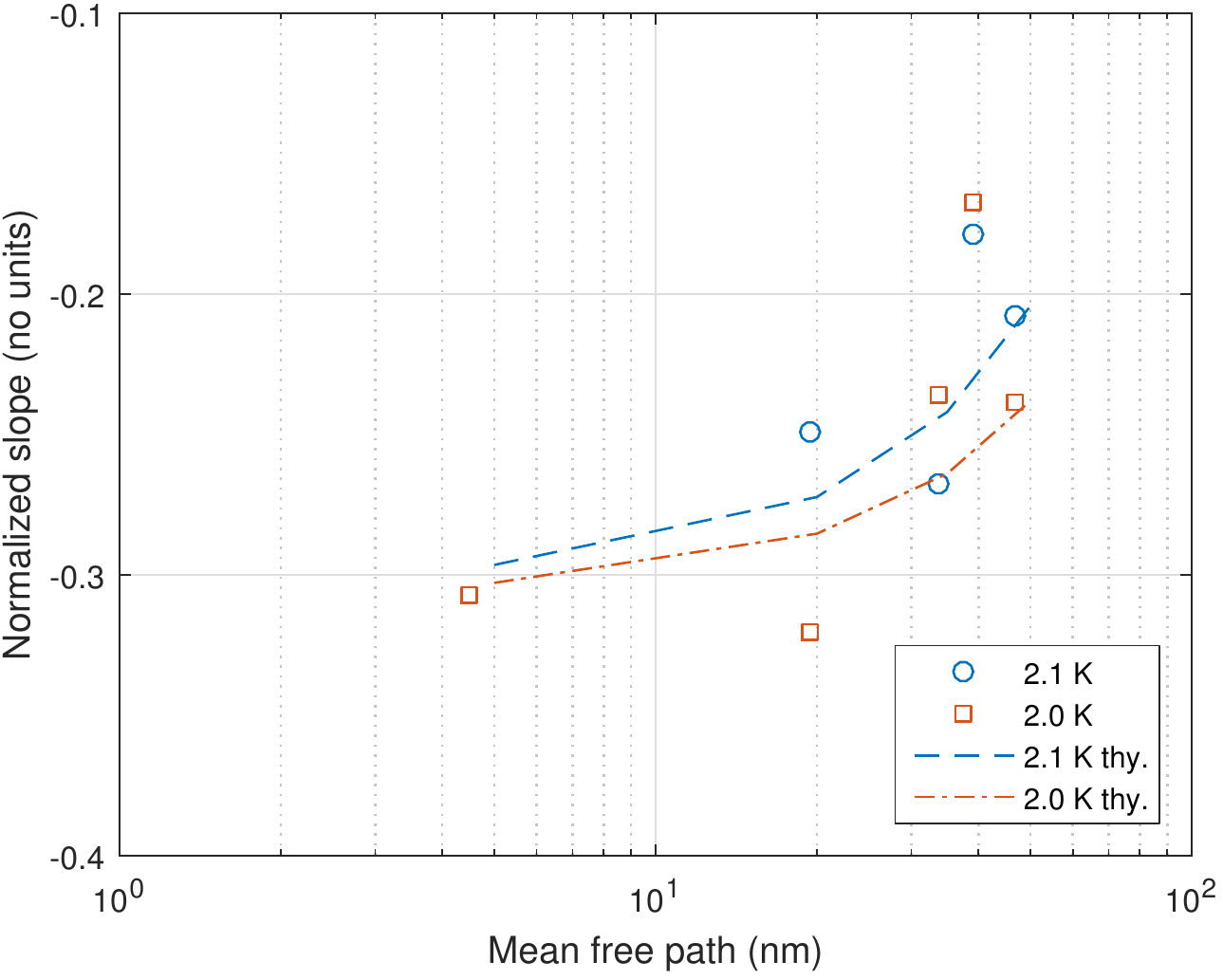}
    \caption{Normalized logarithmic slopes of the surface resistance in field, plotted over mean free path. The plot has been limited to two temperatures to reduce visual clutter.\label{fig:log_slope_mfp}}
    \end{subfigure}
\end{widepage}
\caption{Normalized logarithmic slopes of the BCS surface resistance vs. temperature and mean free path.}
\end{figure}

Isolating representative results from one cavity, in Fig. \ref{fig:te13_log}, we see that both the theory of \cite{gurevich2014} and experimental results approximately show this logarithmic behavior. Normalized by the low-field BCS surface resistance ($\Gamma = \nicefrac{X}{R_\text{BCS,0}}$), this slope shows a wide variation across cavities\footnote{This is to be expected for the wide variation in $\ell$ across the cavities.} but generally exhibits a weak increase with temperature; that is, as the temperature increases, the normalized slope becomes less negative, leading to a weaker anti-Q-slope; this holds as well for cavities approaching the clean limit. Fig. \ref{fig:log_slope_normalized} shows this behavior. The variation in slope is approximately bounded on the lower end by the normalized logarithmic slope for minimal overheating, {\em i.e.} $\alpha \approx 0$, indicating once more that overheating plays a role in the magnitude of the anti-Q-slope. It is noteworthy here that this slight increase in slope with temperature is reproduced by the model: for fixed $\alpha'$ for a given cavity, the increase in the low-field $R_\text{BCS}$ with temperature brings about a corresponding increase in the original, non-normalized overheating parameter $\alpha$; higher $\alpha$, in turn, corresponds with more overheating and thus a weaker anti-Q-slope. These results serve to confirm the original theory provided in \cite{gurevich2014} and our modifications here.

Studying this same slope as a function of mean free path by combining data from all cavities at each temperature, we find that the logarithmic slope becomes less negative as the mean free path increases, as shown in Fig. \ref{fig:log_slope_mfp}. Also shown in Fig. \ref{fig:log_slope_mfp} are the theoretical predictions based on Eqs. \ref{eq:g13} and \ref{eq:g14} with the mean-free-path-dependent $\alpha'$, corresponding well with experimental observations. This is a strong indication in support of the conclusions drawn in the previous section regarding the relationship between the mean free path and the anti-Q-slope: longer mean free paths correspond directly with stronger overheating and thus less-negative normalized logarithmic slopes in $R_\text{BCS}$ vs. $B$.

\section*{Flux Trapping and Optimal Doping Level}

\subsection*{Measuring Sensitivity to Trapped Flux}

As mentioned in the introduction, nitrogen-doped niobium SRF cavities are susceptible to large increases in the residual resistance $R_0$ due to trapped magnetic flux. While the reduction in BCS resistance described above is very enticing, any increase in $R_0$ can counteract this, leading to decreased performance in an SRF accelerating structure.

\begin{figure}[p]
\begin{widepage}
    \begin{subfigure}[t]{0.48\textwidth}
    \centering
    \includegraphics[width=\linewidth]{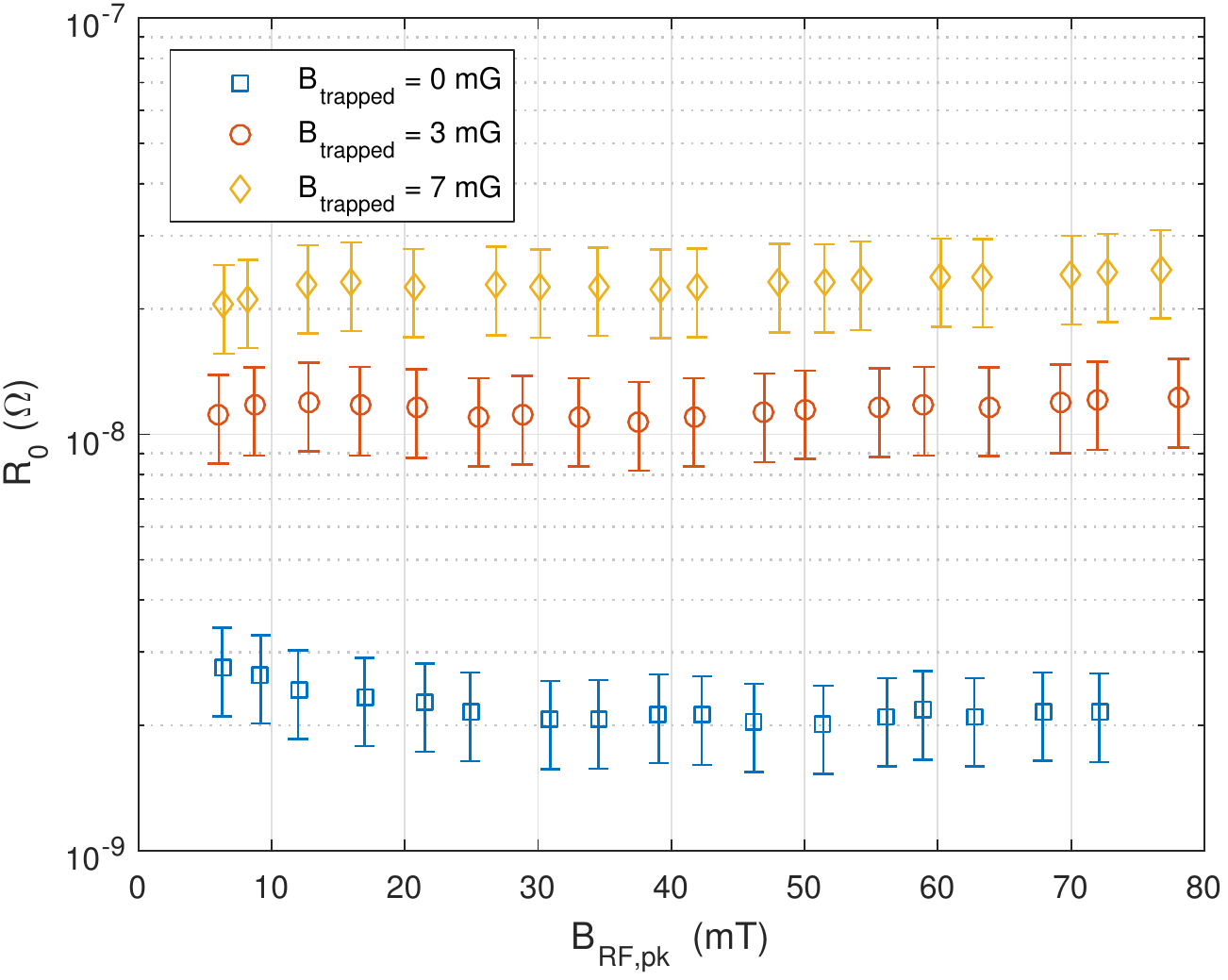}
    \caption{Representative $R_\text{0}$ vs. $B_\text{RF,pk}$ for a 1.3 GHz TESLA cavity, at varying levels of trapped magnetic flux. \label{fig:r0_vs_e}}
    \end{subfigure}\hfill
    \begin{subfigure}[t]{0.48\textwidth}
    \centering
    \includegraphics[width=\linewidth]{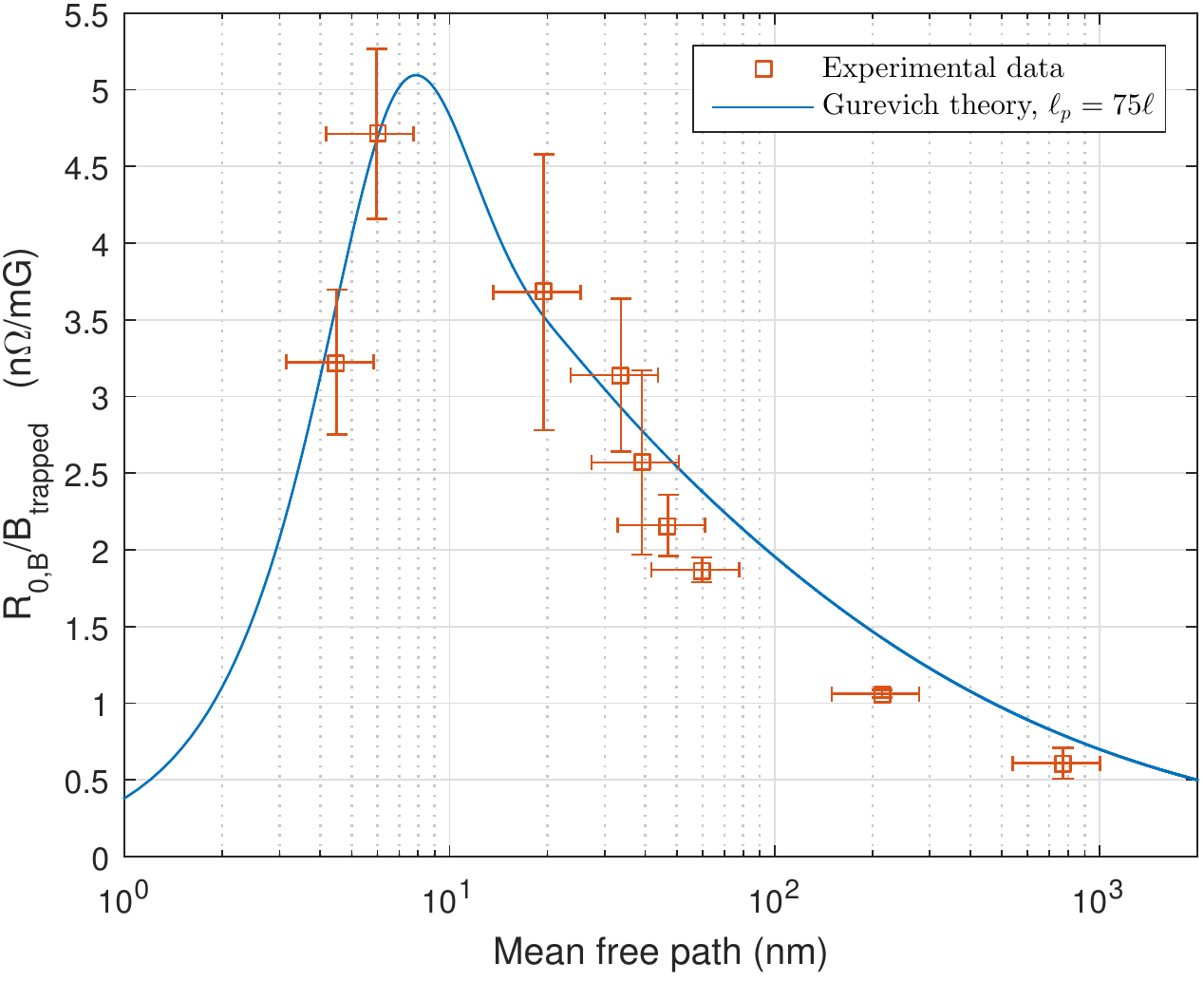}
    \caption{Experimental data and theoretical prediction for sensitivity of residual resistance to trapped magnetic flux as a function of mean free path. \label{fig:sensitivity_vs_mfp}}
    \end{subfigure}
\end{widepage}
\end{figure}

To study the effects of trapped magnetic flux on nitrogen-doped cavities, we tested many 1.3~GHz single-cell TESLA cavities at varying doping level\footnote{These cavities and preparations were the same as those used for the study on the field-dependent BCS surface resistance.}, measured by the mean free path $\ell$, with varying cool-down rates and externally-applied uniform DC magnetic fields in order to study the residual resistance in each cavity as a function of trapped magnetic flux\cite{gonnellaJAP}. Trapped flux was measured by flux gate magnetometers placed on the outer surface of the cavity at the iris. Figure \ref{fig:r0_vs_e} shows representative results for the residual resistance as a function of accelerating field for several amounts of trapped flux.

The residual resistance was found to increase linearly with trapped flux, leading to the simple scaling relation given in Eq. \ref{eq:flux}, where $R_\text{0,B}$ is the portion of the residual resistance due to trapped flux losses:

\begin{equation}
S = \frac{R_\text{0,B}}{B_\text{trapped}} \label{eq:flux}
\end{equation}

Here, $S$ provides a convenient measure of the sensitivity of the residual resistance to trapped flux for a given cavity surface. When analyzing $S$ as a function of mean free path, we found that our results compared agreeably with theoretical predictions\cite{gc13,gonnellaJAP}. The theory involves RF losses due to the oscillation of magnetic flux vortices in the superconductor, with the sensitivity varying with mean free path, and with a fit parameter relating the mean free path to the mean spacing between flux pinning sites. From our cavity data we fitted a mean pinning site spacing $\ell_\text{p}=75\ell$, as shown in Fig. \ref{fig:sensitivity_vs_mfp}\cite{gonnellaJAP}. 

\subsection*{Optimizing the Mean Free Path}

\begin{figure}[p]
\centering
\includegraphics[height=0.5\textwidth]{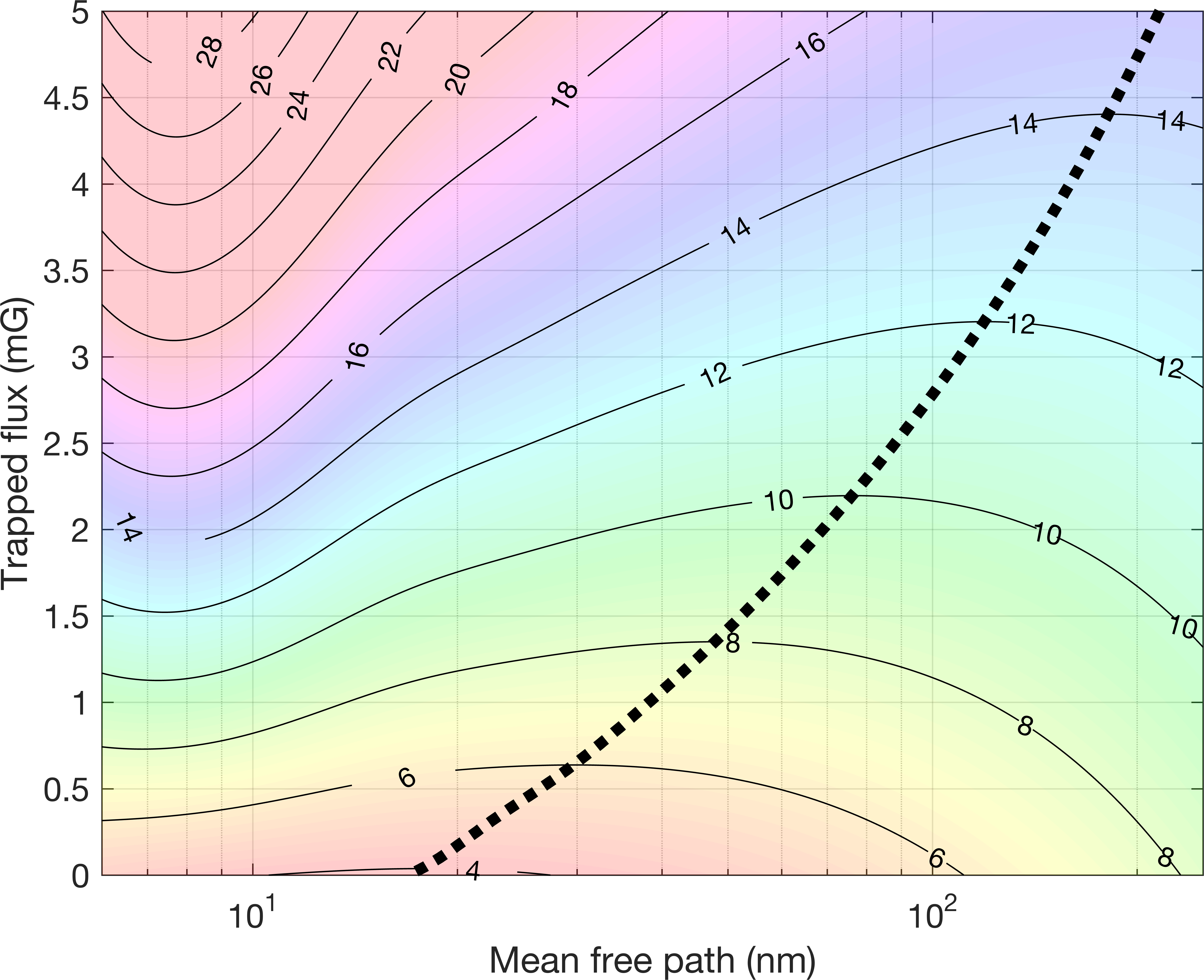}
\caption{Total surface resistance $R_\text{s} = R_\text{BCS}(\ell, 2~\text{K}, 16~\text{MV/m}) + R_0(\ell, B_\text{trapped})$ over mean free path and trapped flux. Contour labels are given in n$\Omega$. Heavy dashed line indicates optimal mean free path for a given amount of trapped flux. \label{fig:optimal_doping}}
\end{figure}

Combining this theoretical prediction of trapped flux sensitivity with the prediction of $R_\text{BCS}$ at a typical operating gradient of 16~MV/m and operating temperature of 2~K, we can derive a model to find the optimal mean free path for a given amount of trapped magnetic flux. Adding $R_\text{BCS}(\ell, 2~\text{K}, 16~\text{MV/m}) + R_0(\ell,B_\text{trapped})$ and plotting over mean free path and trapped flux gives the result seen in Fig. \ref{fig:optimal_doping}. Minimizing $R_\text{s}$ by optimizing\footnote{It is important to note here that there exists a second minimum at lower mean free path, below 5~nm; however, in this heavily-doped region, cavities exhibit a very low average quench field. This gradient limitation makes cavities in this regime unusable for accelerator operations.} mean free path for a given amount of trapped flux results in the dashed line shown in the figure.

In the context of accelerator design, this means that one can choose a minimum achievable trapped flux for the accelerating structure and from this determine the ideal mean free path for doping. Any cavities with smaller values of trapped flux will have a lower overall surface resistance, since $R_0(\ell,B_\text{trapped})$ increases monotonically as a function of trapped flux. The amount of trapped flux depends on ambient fields, the quality of the magnetic shielding, and cooldown procedures. With good shielding, typical ambient fields in SRF cryomodules are near 5~mG. For an operating gradient of 16~MV/m at 2~K, the results in Fig. \ref{fig:optimal_doping} show that, unless more than $\sim$50\% of such an ambient field can be expelled during cooldown, light doping is preferable over heavy doping. On the other hand, if the trapped field can be limited to $<2$~mG, then the stronger anti-Q-slope of heavy doping makes it a preferable option.

\section*{Conclusion}

We have found strong evidence that the electron mean free path plays an important role in the surface resistance of SRF materials. Beyond the well-known minimization of the low-field BCS resistance near $\ell=\xi_0/2$, this role is twofold: decreasing the mean free path into the dirty limit causes an increasingly strong anti-Q-slope, resulting in very low BCS resistances at technologically interesting accelerating gradients, but it also increases susceptibility to increases in the effective residual resistance due to trapped magnetic flux. We have drawn a theoretical connection between the observed dependence on mean free path of the field-dependent BCS resistance and the overheating of quasiparticles, which increases linearly with mean free path in the dirty limit. This is linked either to a decreasing energy transfer rate from quasiparticles to phonons or to an additional energy loss channel from quasiparticles due to elastic scattering on impurities which decreases as the mean free path increases. We have used this model to calculate the BCS resistance for short-mean-free-path cavities at operating accelerator gradients, and combined this with theoretical results for the flux-trapping sensitivity to find an optimal mean free path for a given technologically achievable trapped flux.

\section*{Acknowledgments}

This work was supported by the U.S. DOE LCLS-II High Q Project and NSF Grant PHY-1416318. We would like to thank A. Gurevich for helpful discussions on both \cite{gurevich2014} and \cite{gc13}. We would also like to thank A. Grassellino for bringing nitrogen doping to the field of SRF.

\bibliographystyle{abbrv}
\bibliography{paper}

\end{document}